\documentclass[aps,twocolumn,nobibnotes,pra]{revtex4}
\usepackage{amsmath,epsfig,graphicx,amssymb}
\usepackage{bm}%
\expandafter\ifx\csname package@font\endcsname\relax\else
 \expandafter\expandafter
 \expandafter\usepackage
 \expandafter\expandafter
 \expandafter{\csname package@font\endcsname}%
\fi

\begin{document}
\title{Adiabatic transport of Bose-Einstein condensate in double-well trap}
\author{V.O. Nesterenko$^1$, A.N. Novikov$^{1}$, A.Yu. Cherny$^{1}$,
F.F. de Souza Cruz$^2$, and E. Suraud$^3$}
\affiliation{$^{1}$ Bogoliubov
Laboratory of Theoretical Physics, Joint Institute for Nuclear Research, Dubna,
Moscow region, 141980, Russia}
\affiliation{$^{2}$ Departamento de Fisica, Universidade Federal de
Santa Catarina, Florianopolis, SC, 88040-900, Brasil}
\affiliation{$^{3}$ Laboratoire de Physique Thoretique,
 Universit{\'e} Paul Sabatier,
 118 Route de Narbonne, 31062 cedex,  Toulouse, France}

\date{\today}

\begin{abstract}
A complete irreversible adiabatic transport of Bose-Einstein condensate
(BEC) in a double-well trap is investigated within the mean field
approximation. The transfer is driven by time-dependent (Gaussian) coupling
between the wells and their relative detuning. The protocol successfully
works in a wide range of both repulsive and attractive BEC interaction.
The nonlinear effects caused by the interaction can be turned from detrimental
into favorable for the transport. The results are compared with familiar
Landau-Zener scenarios using the constant coupling. It is shown that the pulsed
Gaussian coupling provides a new transport regime where coupling edges
are decisive and convenient switch of the transport is possible.
\end{abstract}

\maketitle

\section{Introduction}
Nowadays the trapped Bose-Einstein condensate (BEC) is widely recognized as a
source of new fascinating physics and remarkable cross-over with other areas
\cite{Pit_Str_03,Petrick_Smith_02,Dalfovo_99,Leggett_01, Yuk_rew_01,
crossover,Yuk_JPA_96}. Between diverse directions of this field, a large
attention is paid to dynamics of weakly bound condensates (or multi-component
BEC) with emphasis to nonlinear effects caused by interaction between BEC
atoms. The studies cover different aspects of boson Josephson junctions
\cite{Yuk_JPA_96,Smerzi_PRL_97,Raghavan_PRA_99,
Albiez_exp_PRL_05,Zang_PRA_08,Nista_08}, population of topological states
\cite{Will_PRA_00,Yuk_04}, tunneling processes
\cite{Wu_PRA_00,Zobay_PRA_00,Liu_PRA_02,
Jona_PRL_03,Witthaut_PRA_06,Morsch_RMP_06,Liu_PRA_06}, transport of condensate
\cite{Nista_08,Weiss_PRA_05_LPL_06,
Graefe_PRA_06,Ne_Bars_08,Ne_lanl_08,Ne_LP_09,Rab_PRA_08,Opat_PRA_09,Liu_PRA_08}
and other topics. Both multi-level or multi-well systems are considered. In the
first case, BEC is confined in a single-well trap and  contains atoms in a few
hyperfine levels, thus forming BEC components. The laser light can couple the
components and initiate various regimes of the population transfer. In the
second case, BEC atoms occupy the same hyperfine state but the trap is
separated by laser-produced barriers into a series of weakly bound wells. The
atoms can tunnel through the barriers and exhibit similar effects as in a
multi-level system. In this case BEC components are represented by populations
of the wells.

  The present work is devoted to complete irreversible transport of BEC in a two-well
trap. We will consider the process when BEC, being initially in
one of the wells, is then completely transferred in a controllable
way the target well and kept there. Such transport can be realized
in multi-well traps \cite{Albiez_exp_PRL_05} and arrays
of selectively addressable traps \cite{arrays}. Being useful for a general
manipulation of the condensate, the irreversible transport can also open
intriguing prospects for producing new dynamical regimes \cite{Ne_LP_09},
investigation of topological states \cite{Will_PRA_00,Yuk_04},
generation of various geometric phases \cite{Ne_LP_09,Balakrishnan_EPJD_06}.
The latter is especially important since geometric phases are considered as
promising information carriers in quantum computing
\cite{Nayak_RMP_08_gp,Zhu_PRL_03_gp,Feng_PRA_07_gp}.

In general the BEC transport can be designed by various ways: Rabi switch
\cite{Nista_08}, time-dependent potential modulation \cite{Weiss_PRA_05_LPL_06},
Rosen-Zener (RZ) method \cite{Liu_PRA_08}, adiabatic population transfer
\cite{Graefe_PRA_06,Ne_Bars_08,Ne_lanl_08,
Ne_LP_09,Rab_PRA_08,Opat_PRA_09,Liu_PRA_08}.
Adiabatic methods seem to be especially effective for the transport
because of their robustness to modest variations of the process parameters
\cite{Vitanov,Brandes_PR_05,kral_RMP_07}. In particular, the Stimulated
Rapid Adiabatic Passage (STIRAP) \cite{Berg_STIRAP} is widely used
for this aim \cite{Graefe_PRA_06,Ne_Bars_08,
Ne_lanl_08,Ne_LP_09,Rab_PRA_08,Opat_PRA_09}. This method promises
robust and complete population transfer of the
ideal condensate but suffers from nonlinearity effects
caused  by interaction between BEC atoms.
Then a natural question arises,
is it possible to turn the nonlinearity from detrimental to
favorable factor of adiabatic transport?

In this respect the study of nonlinear Landau-Zener (LZ) tunneling of
BEC in the systems like two-band accelerated optical lattices give a useful and
promising message \cite{Zobay_PRA_00,Jona_PRL_03,Morsch_RMP_06}. It was shown
that in such systems the tunneling is asymmetric, i.e. it can be considerably
enhanced or suppressed by the nonlinearity, depending on its sign.
This means that the LZ scheme \cite{LZ} can serve as a good basis for
nonlinear transport of BEC.

The present paper is devoted to realization of this idea for
adiabatic transport of interacting BEC between two potential
wells. In addition to the familiar LZ case with a constant
coupling $\Omega$ between the wells \cite{LZ}, we will inspect the nonlinear
transport with time-dependent coupling $\Omega (t)$ of a Gaussian form.
Actually this a generalization of both LZ (constant coupling
and linear bias of diabatic energies) \cite{LZ} and RZ (time-dependent
coupling and constant diabatic energies) \cite{RZ_PR_32} methods.
Such generalization is partly motivated by Stark Chirped Rapid Adiabatic
Passage (SCRAP) method \cite{Yatsenko_PRA_99}.
Being based on  LZ protocol with a pulsed laser coupling,
SCRAP turned out to be quite effective in atomic physics.
Note also that a constant coupling is somewhat artificial
and its time-dependent version is more realistic.

As shown below, the nonlinear transfers with constant and time-dependent
couplings have much in common (e.g. asymmetric impact of nonlinearity,
determined by the interaction sign) and both them can
result in effective adiabatic transport at strong nonlinearity.
However the time-dependent (pulsed) protocol is more flexible and
has interesting peculiarities. For example, we get a new transport
regime where the coupling edges play a decisive role. Besides,
the pulsed coupling allows us to switch on/off the transport in much wider
interval of the detuning rate $\alpha$ and control the switch value
$\alpha_s$ through the coupling width $\Gamma$.

In the present paper these peculiarities are studied by using the
stationary spectra, in particular the loops arising there
due to the nonlinearity. Although the
transport problem is not stationary and we actually deal with
time-dependent Gross-Pitaevskii equation \cite{GPE},
such analysis is known to be very instructive
\cite{Wu_PRA_00,Zobay_PRA_00,Liu_PRA_02,Morsch_RMP_06}.
It gives a clear intuitive treatment of quite complicated processes
and allows for getting simple analytical estimations.

The paper is outlined as follows. In Sec. 2 the mean-field formalism is
outlined. In Sec. 3 we present the transport protocol.
In Sec. 4 results of the calculations are discussed.
The summary is done in Sec. 5.

\section{Model}
\subsection{General formalism}

The calculations have been performed in the mean-field approximation by using
the nonlinear Schr\"odinger, or Gross-Pitaevskii equation \cite{GPE}
\begin{equation}
\label{eq:NLSE}
i\hbar{\dot \Psi}({\vec r},t) = [-\frac{\hbar^2}{2m}\nabla^2
+ V_{ext}({\vec r},t)
+ g_0|\Psi({\vec r},t)|^2]\Psi({\vec r},t)
\end{equation}
where the dot means the time derivative, $\Psi({\vec r},t)$ is the classical
order parameter of the system, $V_{ext}({\vec r},t)$ is the external trap
potential involving both (generally time-dependent) confinement and coupling,
$g_0=4\pi a/m$ is the parameter of interaction
between BEC atoms, $a$ is the scattering length and $m$ is the atomic
mass.

The order parameter of BEC in a double-well trap can be expanded as
\cite{Smerzi_PRL_97}
\begin{equation}\label{eq:Psi}
\Psi({\vec r},t)=\sqrt{N}(\psi_1(t)\Phi_1({\vec r})+\psi_2(t)\Phi_2({\vec r}))
\end{equation}
where $\Phi_k({\vec r})$ is the static ground state solution of
(\ref{eq:NLSE}) for the isolated k-th well ($k=1,2$)
and
\begin{equation}\label{order_par}
\psi_k(t)=\sqrt{N_k(t)}e^{i\phi_k(t)}
\end{equation}
is the corresponding amplitude expressed via the relative population
$N_k(t)$ and phase $\phi_k(t)$. The total number of atoms $N$
is fixed
\begin{equation}
\int d\vec r |\Psi({\vec r},t)|^2/N=N_1(t)+N_2(t) = 1 \; .
\end{equation}

Substituting (\ref{eq:Psi}) to (\ref{eq:NLSE}) and integrating out the spatial
distributions $\Phi_k({\vec r})$
we get
\cite{Smerzi_PRL_97}
\begin{eqnarray}\label{psi_1(t)}
  i{\dot \psi}_1 &=& [E_1(t)+ UN|\psi_1|^2]\psi_1
- \Omega(t) \psi_2 \; ,
\\
\label{psi_2(t)}
 i{\dot \psi}_2 &=& [E_2(t)+ UN|\psi_2|^2]\psi_2
- \Omega(t) \psi_1
\end{eqnarray}
where
\begin{equation}\label{Om}
 \Omega (t) = - \frac{1}{\hbar}\int d{\vec r} \;
 [\frac{\hbar^2}{2m}\nabla\Phi^*_1 \cdot\nabla\Phi_2
 +\Phi^*_2 V_{ext}(t)\Phi_1]
\end{equation}
is the coupling between BEC fractions
\begin{equation}\label{E}
  E_k(t)= \frac{1}{\hbar}\int d{\vec r} \;
  [\frac{\hbar^2}{2m}|\nabla\Phi^*_k|^2
  +\Phi^*_k V_{ext}(t)\Phi_k]
\end{equation}
is the depth of the k-th well, and
\begin{equation}\label{U}
  U_k= \frac{g_0}{\hbar}\int d{\vec r} \; |\Phi_k|^4 \;
\end{equation}
labels the interaction between BEC atoms in the same well.
Here we assume that $U_1=U_2=U$. The values $\Omega (t)$, $E_k(t)$,
and $U$ have the  dimension of frequency.

In the present study we use a Gaussian coupling
with a common amplitude $K$
\begin{equation}\label{Omega}
\Omega (t)=K {\bar\Omega}(t), \quad
\bar{\Omega}(t)=\exp\{-\frac{(\bar{t}-t)^2}{2\Gamma^2}\}
\end{equation}
where $\bar{t}$ and $\Gamma$ are centroid and width parameters.
Then scaling
(\ref{psi_1(t)})-(\ref{psi_2(t)}) by $1/(2K)$ gives
\begin{eqnarray}
\label{spsi_1(t)}
  i\dot{\psi}_1 &=& [\bar{E}_1(t)
  +\Lambda |\psi_1|^2]\psi_1
- \frac{1}{2} \bar{\Omega}(t) \psi_2
\\
\label{spsi_2(t)} i\dot{\psi}_2 &=& [\bar{E}_2(t)
  +\Lambda |\psi_2|^2]\psi_2 - \frac{1}{2} \bar{\Omega}(t) \psi_1
\end{eqnarray}
where
\begin{equation}\label{Lambda}
\bar{E}_k(t)=\frac{E_k(t)}{2K}, \quad \Lambda=\frac{UN}{2K}
\end{equation}
and the scaled time, $2Kt \to t$, is dimensionless. In
(\ref{spsi_1(t)})-(\ref{spsi_2(t)}), the key parameter $\Lambda$ is responsible
for the interplay between the coupling and interaction.

Substituting (\ref{order_par}) into (\ref{spsi_1(t)})-(\ref{spsi_2(t)}) leads
to the system of equations for the populations $N_{k}(t)$ and phases
$\phi_{k}(t)$:
\begin{eqnarray}\label{eq:N_dot}
  {\dot N}_k&=& - \bar{\Omega}(t)\sqrt{N_j N_k}
 \sin (\phi_j-\phi_k) \; ,
\\
  {\dot \phi}_k &=& -[\bar{E}_k(t) +\Lambda N_k]
+ \frac{1}{2}\bar{\Omega}(t)\sqrt{\frac{N_j}{N_k}}
 \cos (\phi_j-\phi_k)
\label{eq:phi_dot}
\end{eqnarray}
with $j \ne k$. Considering $N_k$ and $-\phi_k$ as conjugate
variables and using the linear canonical transformation \cite{Ne_LP_09}
\begin{eqnarray}\label{z_trans}
z&=&N_1-N_2, \quad, Z=N_1+N_2=1 \; ,
\\
\theta&=&\frac{1}{2}(\phi_2-\phi_1)
\quad, \Theta=-\frac{1}{2}(\phi_1+\phi_2)
\end{eqnarray}
one may extract the integral of motion $Z$ and corresponding total
phase $\Theta$ from equations (\ref{eq:N_dot})-(\ref{eq:phi_dot}) and
thus reduce the problem to a couple of equations for
new unknowns, population imbalance $z$ and phase difference $\theta$.
These equations read
\begin{eqnarray}\label{z_dot}
 \dot z &=& -\bar{\Omega}(t)\sqrt{1-z^2} \sin 2\theta
\\
\dot\theta &=& \frac{1}{2}[\Delta(t)+\Lambda z
+\bar{\Omega}(t)\frac{z}{\sqrt{1-z^2}} \cos 2\theta]
\label{theta_dot}
\end{eqnarray}
where
\begin{equation}\label{detuning}
\Delta (t)=\bar E_1(t) - \bar E_2(t)=\alpha t.
\end{equation}
is the detuning (difference in well depths).
In accordance with LZ practice we assume for $\Delta (t)$
a linear time-dependence with the rate $\alpha$.

Equations of motion (\ref{z_dot})-(\ref{theta_dot}) are invariant under
transformations
\begin{equation}\label{symm_1}
\Lambda \to -\Lambda, \quad \alpha \to -\alpha,  \quad  \theta \to -\theta+\frac{\pi}{2}
\end{equation}
or
\begin{equation}\label{symm_2}
\alpha \to -\alpha, \quad z \to -z,  \quad  \theta \to -\theta \; .
\end{equation}
The former relates the transport protocols for repulsive
and attractive BEC in one direction while the latter connects
the transport in opposite directions (with corresponding interchange
of the initial conditions).

Note that equations (\ref{z_dot})-(\ref{theta_dot}) allow a classical analogy
with $z$ and  $\theta$ treated as conjugate variables. It is easy to verify
that these equations can be recast into the canonical form
\begin{equation}\label{eq:can_eq}
 {\dot z} = - \frac{\partial {\rm H}_{cl}}{\partial \theta} \; ,
 \quad
{\dot \theta} =  \frac{\partial {\rm H}_{cl}}{\partial z}
\end{equation}
with the classical Hamiltonian
\begin{equation}\label{eq:class_ham}
 {\rm H}_{cl}= \frac{1}{2}[\Delta(t) z + \frac{\Lambda}{2}z^2
- \bar{\Omega}(t)\sqrt{1-z^2}\cos 2\theta]
\end{equation}
and chemical potential \cite{Lipp_book}
\begin{equation}\label{chem_pot}
 \mu = {\rm H}_{cl}+V_{int} =\frac{1}{2}[\Delta(t) z+\Lambda z^2
- \bar{\Omega}(t)\sqrt{1-z^2}\cos 2\theta ]
\end{equation}
where $V_{int}=\Lambda z^2/4$.

It is easy to see that, up to notation, Eqs.
(\ref{z_dot}), (\ref{theta_dot}) and (\ref{eq:class_ham})
coincide with those in \cite{Smerzi_PRL_97}. However,
unlike the previous studies  of the {\it oscillating} BEC fluxes in traps with
constant parameters \cite{Smerzi_PRL_97,Nista_08}, we will
deal with {\it irreversible} BEC transport by monitoring time-dependent
parameters $\bar E_k(t)$ and $\bar\Omega (t)$.

\subsection{Stationary states}

Since we are interested in adiabatic (very slow)
transport, the analysis of stationary states can be useful.
These states are defined by the condition
\begin{equation}\label{eq:can_eq}
 {\dot z} = {\dot \theta} = 0 \; .
\end{equation}
Despite the fact that our Hamiltonian and thus the variables $z$
and $\theta$ actually depend on time, this dependence is assumed
to be slow enough to make (\ref{eq:can_eq}) relevant.

Under condition (\ref{eq:can_eq}) equations
(\ref{z_dot})-(\ref{theta_dot}) give
\begin{eqnarray}\label{z_dot_stat_2}
&& \theta = \frac{\pi}{2}n \; ,
\\
&& \Delta(t)+z(\Lambda
\pm\frac{\bar{\Omega}(t)}{\sqrt{1-z^2}}) = 0
\label{theta_dot_stat_2}
\end{eqnarray}
where $n$ is an integer number. Equation (\ref{theta_dot_stat_2})
has "$+$" or "$-$" for even and odd $n$, respectively. The analysis of
(\ref{theta_dot_stat_2}) finds two real roots for
$|\Lambda| < \bar{\Omega}$ and four real roots for
$|\Lambda| > \bar{\Omega}$, i.e. at high nonlinearity.

Substitution of the stationary solutions of
(\ref{z_dot_stat_2})-(\ref{theta_dot_stat_2}) into (\ref{chem_pot}) gives the
chemical potentials $\mu_-(t)$ and $\mu_+(t)$, representing the eigenvalues of
the stationary states, i.e. (\ref{order_par}) casts into
$\psi_k(t)=\sqrt{N_k(t)}e^{-i\mu t}$. In the four-root case, the repulsive
interaction ($\Lambda >0$) leads to one solution for $\mu_+(t)$ and three
solutions for $\mu_-(t)$, and vice versa for attractive interaction ($\Lambda
<0$).

As shown below, three roots form a nonlinear loop in the stationary spectra.
For $|\Lambda| > \bar\Omega$, the time length of the loop reads
\cite{Liu_PRA_02}
\begin{equation}\label{Delta_c}
 t_c = \frac{(|\Lambda|^{2/3}-(\bar\Omega(t_c))^{2/3})^{3/2}}{\alpha}
 \; ,
\end{equation}
where $\bar\Omega (t_c) \to 1$ for the constant coupling.
It is seen that the length increases with magnitude of nonlinearity $\Lambda$
and decreases with detuning rate $\alpha$. The nonlinear structures (single loops,
double loops, butterfly, ...)  can appear in the stationary spectra
for smaller nonlinearity $|\Lambda| < \bar\Omega$ as well, see discussion
in Ref. \cite{ne_LP_10}.

\section{Transport protocol}

In the standard LZ protocol without nonlinearity \cite{LZ}, the final probabilities
of two competing processes, Landau-Zener tunneling between adiabatic levels ($P_{LZ}$)
and adiabatic following ($P$),  read
\begin{equation}\label{LZ}
  P_{LZ}=e^{-\frac{\pi \Omega^2}{2\alpha}}, \quad P=1-P_{LZ} \; .
\end{equation}
Both processes are controlled by a {\it constant} coupling $\Omega$ and
detuning rate $\alpha$. A complete adiabatic transfer $P = 1$ takes place in the
adiabatic limit $\alpha \to 0$. As  shown in \cite{Wu_PRA_00,Zobay_PRA_00,Liu_PRA_02},
the nonlinearity caused by interaction between BEC atoms essentially
modifies the transfer (\ref{LZ}). Namely, the diabatic LZ  tunneling
becomes possible even at $\alpha \to 0$. Moreover, the impact of nonlinearity
is asymmetric, i.e. it favors or suppresses the LZ tunneling and inversely affects
the adiabatic following, depending on the interaction sign \cite{Zobay_PRA_00}.
This suggests a principle possibility to use the LZ scheme for the complete
adiabatic transport in the nonlinear case.

In the present study, this point is scrutinized  for a constant
and time-dependent Gaussian coupling. The relevant protocols are illustrated
in Fig. 1 where the detuning (\ref{detuning}) is modeled by
\begin{equation}\label{E_evol}
\bar{E}_1(t)=\frac{1}{2}\alpha t , \; \bar{E}_2(t)=-\frac{1}{2}\alpha t
\end{equation}
with $\alpha >0$ and $-t_s < t < t_s$. We assume that at early time $t \le
-t_s$ the diabatic energies are tuned from the symmetry values
$\bar{E}_1=\bar{E}_2=0$ to the values $\bar{E}_1(-t_s)$ and $\bar{E}_2(-t_s)$,
then exhibit the linear  evolution (\ref{E_evol}) at $-t_s < t < t_s$ and
finally come back to the zero symmetry values. In the experimental setup such
manipulations can be produced by varying the well depths. In what follows, we
will present the results only for  $-t_s < t < t_s$.

\begin{figure}
\includegraphics[height=9cm,width=4.0cm,angle=-90]{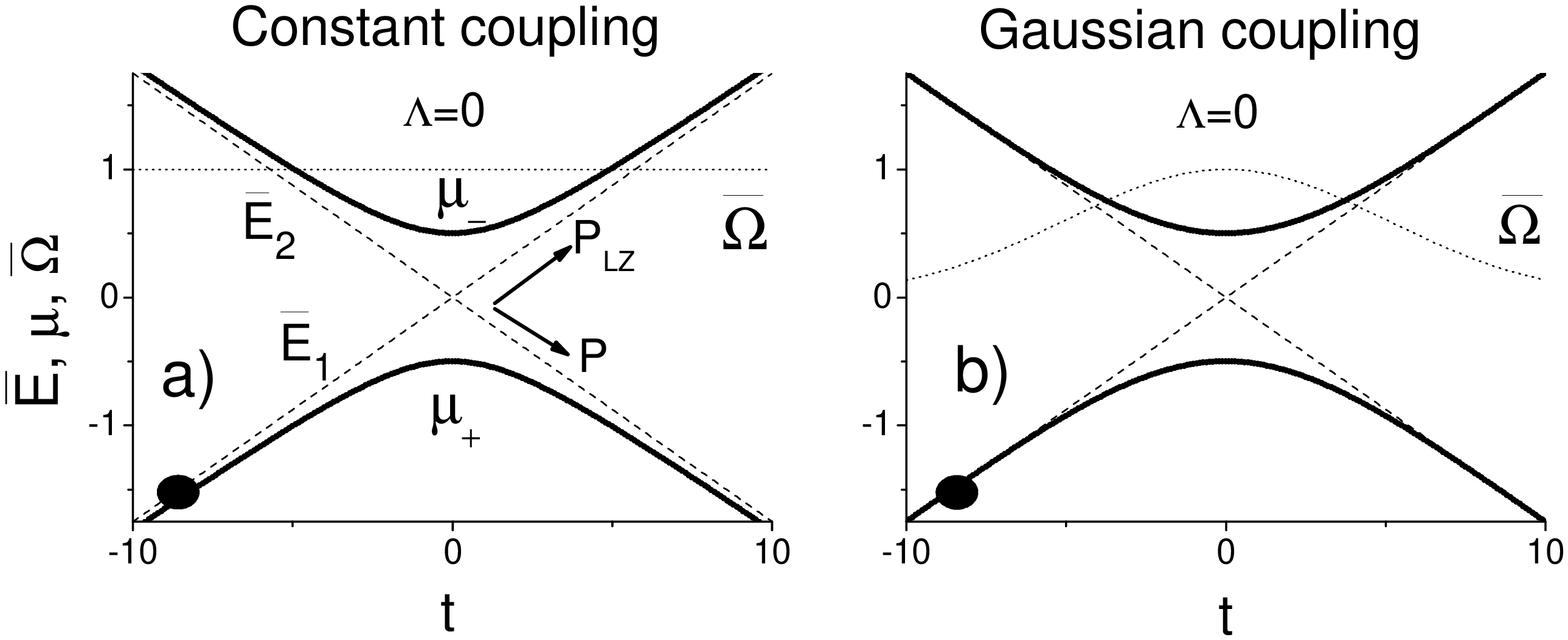}
\caption{ \label{fig:fig1_ill} a): Stationary adiabatic levels $\mu_{\pm}$
(bold curves) calculated in the linear case ($\Lambda$=0) for the constant
coupling $\bar{\Omega}$=1 (dotted line) and $\alpha$=0.35. The diabatic levels
$\bar E_{1,2}$ (strait dash curves) are given for the comparison. BEC is
initially placed at the lower level as indicated by the bold dot. The arrows
show the diabatic $P_{LZ}$ and adiabatic $P$ transfers with the probabilities
(\protect\ref{LZ}). b): The similar scheme for the time-dependent Gaussian
coupling (\protect\ref{Omega}).
}
\end{figure}

The conventional LZ scheme is illustrated in Fig. \ref{fig:fig1_ill}a). Despite
its artificial properties (constant coupling
and infinite diabatic energies at $t \to \pm\infty$), it
is known to give quite realistic transition probabilities (\ref{LZ}).
The reason is that  the LZ transfer actually occurs only during
a finite time of close approaching the diabatic energies near $t \approx 0$,
when $\Delta (t) < \Omega$. The lateral regions of early and late
time, where the coupling impact is negligible, are irrelevant. Hence
the LZ artifacts are not essential.

However, if anyway LZ works only for a finite time, then it is natural to
use a time-dependent coupling of a certain duration. In this connection,
we propose
an effective and flexible transport protocol where a time-dependent coupling
$\bar\Omega(t)$ of a Gaussian form (\ref{Omega}) is used, see Fig. \ref{fig:fig1_ill}b).
Then, in addition to the usual LZ control parameters, detuning rate $\alpha$
and coupling amplitude $K$, we get a new one, the Gaussian width $\Gamma$.
As discussed below, this protocol allows a rapid and complete switch of
adiabatic BEC transport in a wider range of $\alpha$ and provides
a new transport regime where not the center but edges of the coupling
$\bar\Omega(t)$ are important. This makes the new protocol more effective
and flexible. Hereafter it is referred as a Gaussian Landau-Zener (GLZ)
protocol. Obviously, the GLZ is also a generalization of the Rosen-Zener scheme
\cite{RZ_PR_32} which exploits a time-dependent coupling but constant
diabatic energies.

\begin{figure}
\includegraphics[height=8.5cm,width=8cm,angle=-90]{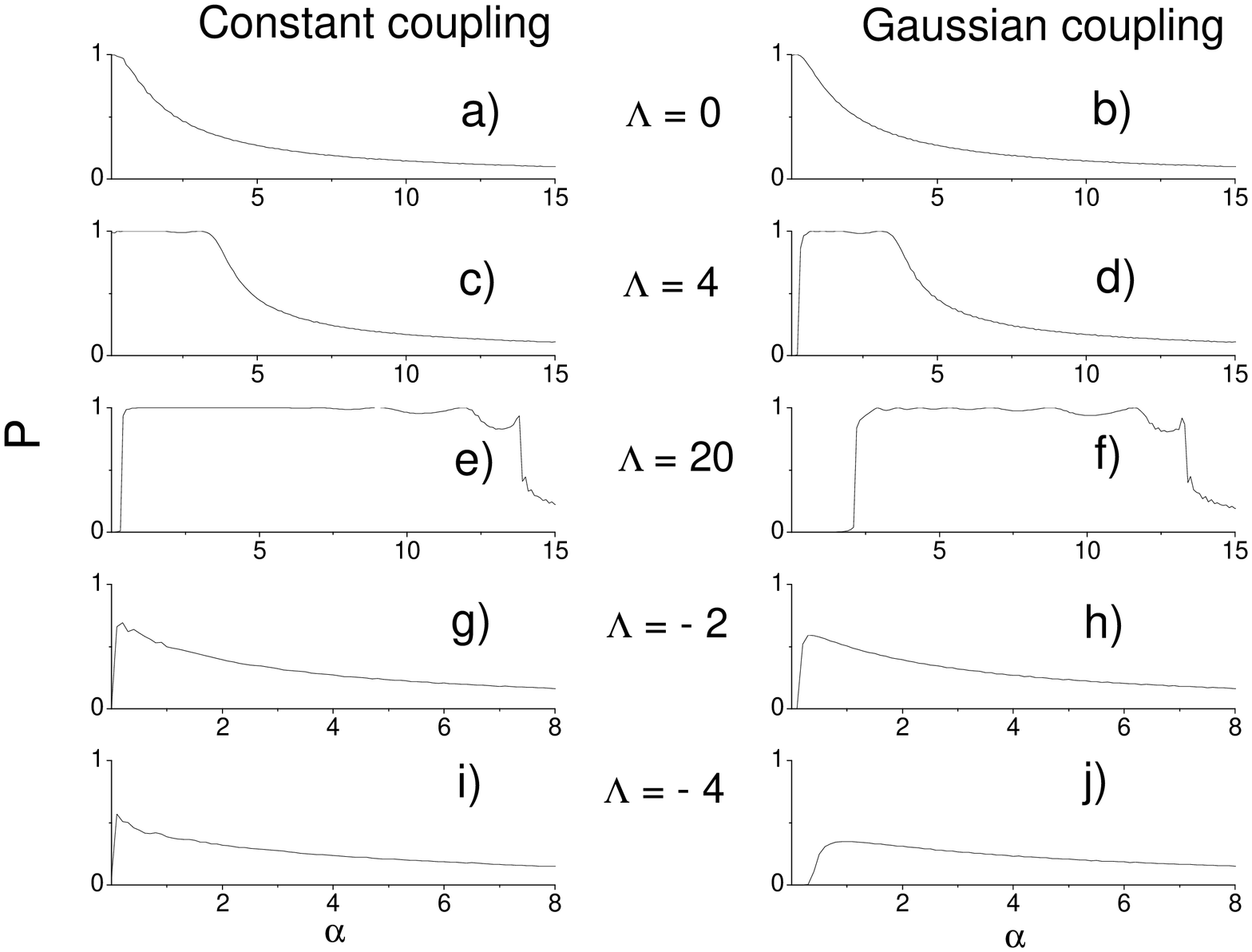}
\caption{
\label{fig:fig2_P}
Dependence of the final adiabatic population $P$ of the 2nd well
on the detuning rate $\alpha$ and nonlinearity $\Lambda$ within
the LZ (left) and GLZ (right) protocols.
}
\end{figure}

In the present calculations we use the Gaussian coupling (\ref{Omega})
with the centroid at $\bar{t}$ =0 and width parameter $\Gamma$=5,
apart from of Fig. 7 where various $\Gamma$ are applied.
The coupling full width at half maximum is
FWHM=2.355 $\Gamma$ = 11.78. The monitoring of the coupling $\bar\Omega(t)$
which is the penetrability of the barrier between the wells, can be
produced in experiment by variation of the spacing between the wells.
For a time-dependent coupling, the adiabatic following takes place
under the condition \cite{Vitanov}
\begin{equation}\label{eq:adia_two_wells}
\frac{1}{2}|\Omega(t)\dot{\Delta}(t)-\dot{\Omega}(t)\Delta(t)|
\ll |\Omega^2(t)+\Delta^2(t)|^{3/2}
\end{equation}
which requires slow rates and large values of the coupling and/or detuning.

Figure 1 shows that in the linear case both LZ and GLZ protocols give very
similar results. The only tiny difference is that GLZ adiabatic energies
$\mu_{\pm}$, calculated with (\ref{chem_pot}), less deviate from
$\bar E_{1,2}$ at early and late time.
In both protocols, the transfer is determined by the region
where $|\mu_- - \mu_+| \approx \bar\Omega(t=0)=1$. As shown below,
much more serious difference between LZ and GLZ  appears
in the nonlinear case.

\section{Results and discussion}

The main results of our calculations are presented in Figs. 2-8.

In Fig. \ref{fig:fig2_P}, the final LZ and GLZ populations of the second
well, $P=N_2(t=+\infty )$, are shown as a function of $\alpha$
for different values of the nonlinear parameter $\Lambda$.
In all the cases, the BEC is initially in the first well, i.e.
$N_1(t=-\infty )=1, N_2(t=-\infty )=0$.
At the first glance, the LZ and GLZ transfers look quite similar.
Without the interaction ($\Lambda$ = 0) the processes are identical.
For the repulsive interaction $\Lambda=4$ and 20, both protocols
produce a complete transfer within a wide $\alpha$-plateau.
The stronger the interaction, the wider the plateau.
At sufficiently high $\alpha$, the process
becomes too rapid and adiabatic transfer naturally fails, leading to
decreasing $P$. In this region, behaviour of P depends on $\Lambda$
but is the same for LZ and GLZ.
\begin{figure}
\includegraphics[height=8.5cm,width=5cm,angle=-90]{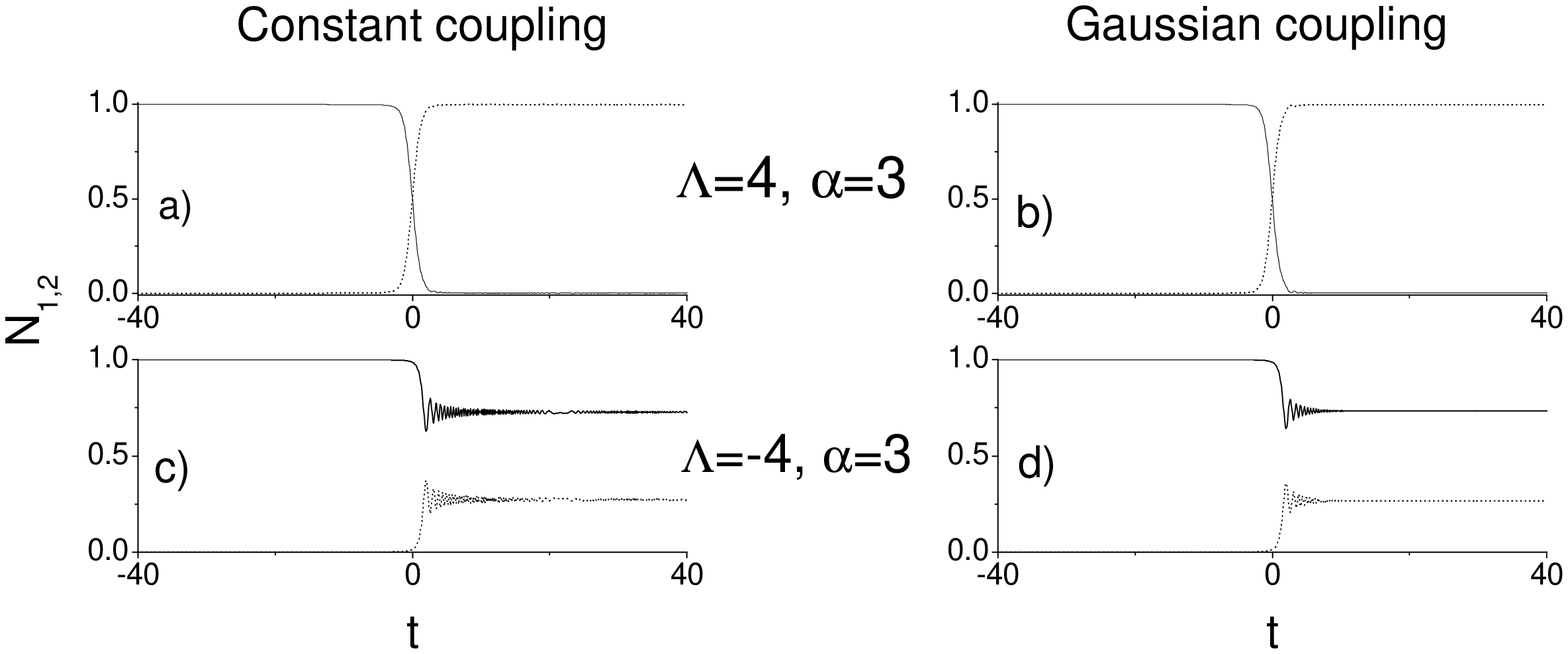}
\caption{ \label{fig:fig3_N1N2} LZ (left) and GLZ (right) time evolution of
populations $N_1$ (solid curve) and $N_2$ (dash curve) of the first and second
wells for the repulsive (a, b) and attractive (c, d) BEC.
}
\end{figure}
\begin{figure}
\includegraphics[height=6cm,width=6cm,angle=-90]{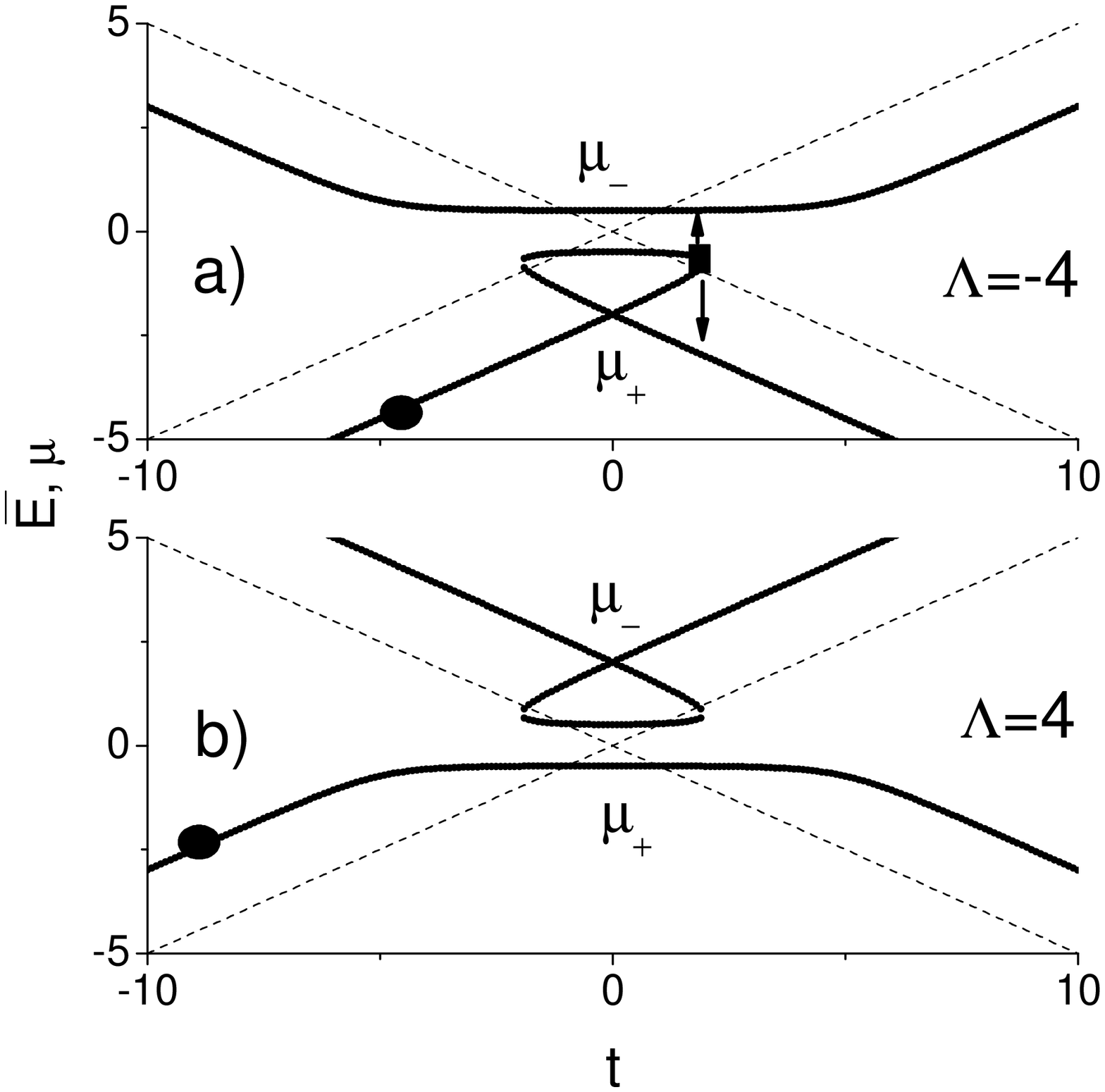}
\vspace{0.4cm} \caption{ \label{fig:fig4_attr} GLZ adiabatic energies
$\mu_{\pm}$ (solid curves) for attractive $\Lambda$=-4 (a) and repulsive
$\Lambda$=4 (b) interaction. The diabatic energies (dash lines) are given for
the convenience of comparison. In both plots the rate $\alpha$=3 is used.
BEC is initially placed at the lower level as depicted by a bold dot.
In the plot a) the terminal loop point and directions of tunneling
to diabatic (arrow up) and adiabatic  (arrow down) paths are marked.
}
\end{figure}

It is remarkable that, unlike the STIRAP case \cite{Ne_LP_09},
the repulsive interaction ($\Lambda > 0$) not spoils but even favors
the transport by forming the $\alpha$-plateau. The robustness and
completeness of the transport is illustrated in Fig. \ref{fig:fig3_N1N2}(a, b)
for the particular case of $\Lambda$=4 and $\alpha$=3.
Instead, following Fig.\ref{fig:fig2_P}(g-j) and Fig. \ref{fig:fig3_N1N2}(c, d),
the attractive interaction ($\Lambda < 0$) damages the adiabatic transport
and the stronger the interaction, the less the final population $P$ \cite{collaps}.
So, in accordance with  \cite{Zobay_PRA_00,Liu_PRA_08}, we obtain
the asymmetric interaction effect when the transfer crucially depends
of the interaction sign.

The origin of this effect is explained in Fig. \ref{fig:fig4_attr} in
terms of  GLZ stationary eigenvalues $\mu_{\pm}$ for the repulsive and
attractive BEC. In both cases BEC is initially placed at
the lower level, i.e. $N_1(t=-\infty)$=1 and $N_2(t=-\infty)$=0.
The nonlinearity causes the loops
in $\mu_{+}$ ($\Lambda$=-4) and  $\mu_{-}$ ($\Lambda$=4) levels.
For $\Lambda$=-4, the loop takes place at the BEC transfer path
and so a part of the condensate arrives to the loop terminal point.
There is no way further and so BEC is enforced to tunnel to
upper or lower levels as indicated by arrows in the plot a). This results in
leaking the population to the upper level $\mu_{-}$ and hence depletion of
the adiabatic transfer. Instead, for $\Lambda$=4,
the transfer path does not meet this obstacle and so is robust.

\begin{figure}
\includegraphics[height=8.5cm,width=8cm,angle=-90]{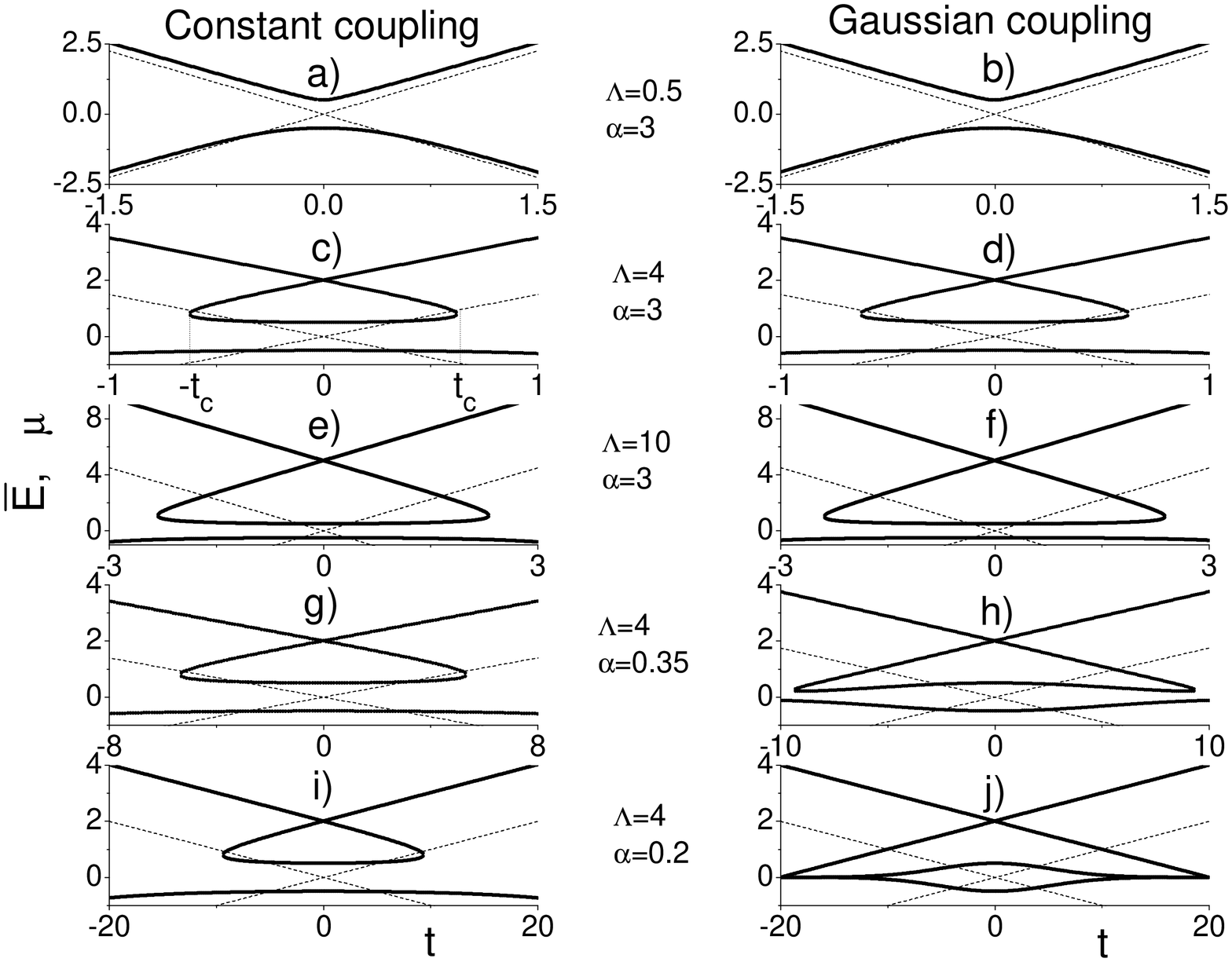}
\vspace{0.4cm}
\caption{ \label{fig:fig5_lam_a_dep} LZ (left) and GLZ (right) chemical
potentials $\mu_{\pm}$ (solid curves) at different nonlinearity $\Lambda$
and detuning rate $\alpha$. The diabatic energies (dash lines) are given
for the convenience of comparison. In the plot c), the loop length
limits $(-t_c, t_c)$ are marked.}
\end{figure}

The previous analysis mainly concerned the common features of LZ and GLZ
protocols. Let's now consider the difference between these two cases, which
takes place at small rates $\alpha$. As seen from Fig. \ref{fig:fig2_P},
for both protocols the nonlinearity results in a window $0 \le \alpha<\alpha_{s}$
where the adiabatic transfer completely vanishes. However, the GLZ window is
much wider than the LZ one, which is a consequence of a finite duration of the
Gaussian coupling.

This point is scrutinized in Fig. \ref{fig:fig5_lam_a_dep} where behaviour of
nonlinear loops in different transport regimes is demonstrated. In the first
regime (plots (a,b)), the loops are absent because of a small nonlinearity
$\Lambda < 1$. The LZ and GLZ stationary spectra and transport features look
similar. In particular, in both cases the adiabatic transport is possible only
in a narrow interval of small rates $\alpha$ \cite{weak_nonlin}. In the second
regime  with a larger nonlinearity $\Lambda > \bar{\Omega}$ (plots (c-f)), the
stationary spectra gain the loops but their length is still less than the
Gaussian width, i.e. $2t_c <$FWHM or $t_c < \Gamma$. Then the LZ and GLZ
spectra remain to be similar and both protocols give a robust adiabatic
transport $P \approx 1$ within common $\alpha$ intervals inside the plateau in
Fig. 2. Note that according to (\ref{Delta_c}), the loop length $2t_c$ rises
with $\Lambda$ and decreases with $\alpha$ as illustrated in Fig. 5. In other
words, the loop behavior is determined not only by the nonlinearity but the
detuning rate as well. The latter becomes decisive in the third regime that
determines the $P \approx 0$ windows at low rates $\alpha$ in Fig. 2. In this
regime (plots (g-j)), the loop becomes comparable or longer than the coupling
width, i.e. $t_c \ge \Gamma$. Then the edges of the Gaussian coupling come to
play and a new regime, which is absent in LZ, arises. At the edges we have
$\bar\Omega(t) \ll 1$. So, in accordance to (\ref{Delta_c}), the  GLZ loop is
extended and becomes longer than the LZ one, see plots (g, h). What is
important for our aims, in this regime the gap  between $\mu_-$ and $\mu_+$
shrinks at the loop edges, thus allowing a partial diabatic population transfer
from $\mu_+$ to $\mu_-$ and corresponding break of the adiabatic following. So,
unlike the LZ, the main effect here takes place not at the coupling maximum but
at its edges. For even longer loops (plot j)), the GLZ leads to merging $\mu_-$
and $\mu_+$ levels at early and late times, which fully breaks the adiabatic
following and causes the windows exhibited in Fig. 2. Conversely, the constant
coupling prevents the  merging and so keeps the adiabatic transfer.
\begin{figure}
\includegraphics[height=7cm,width=4.3cm,angle=-90]{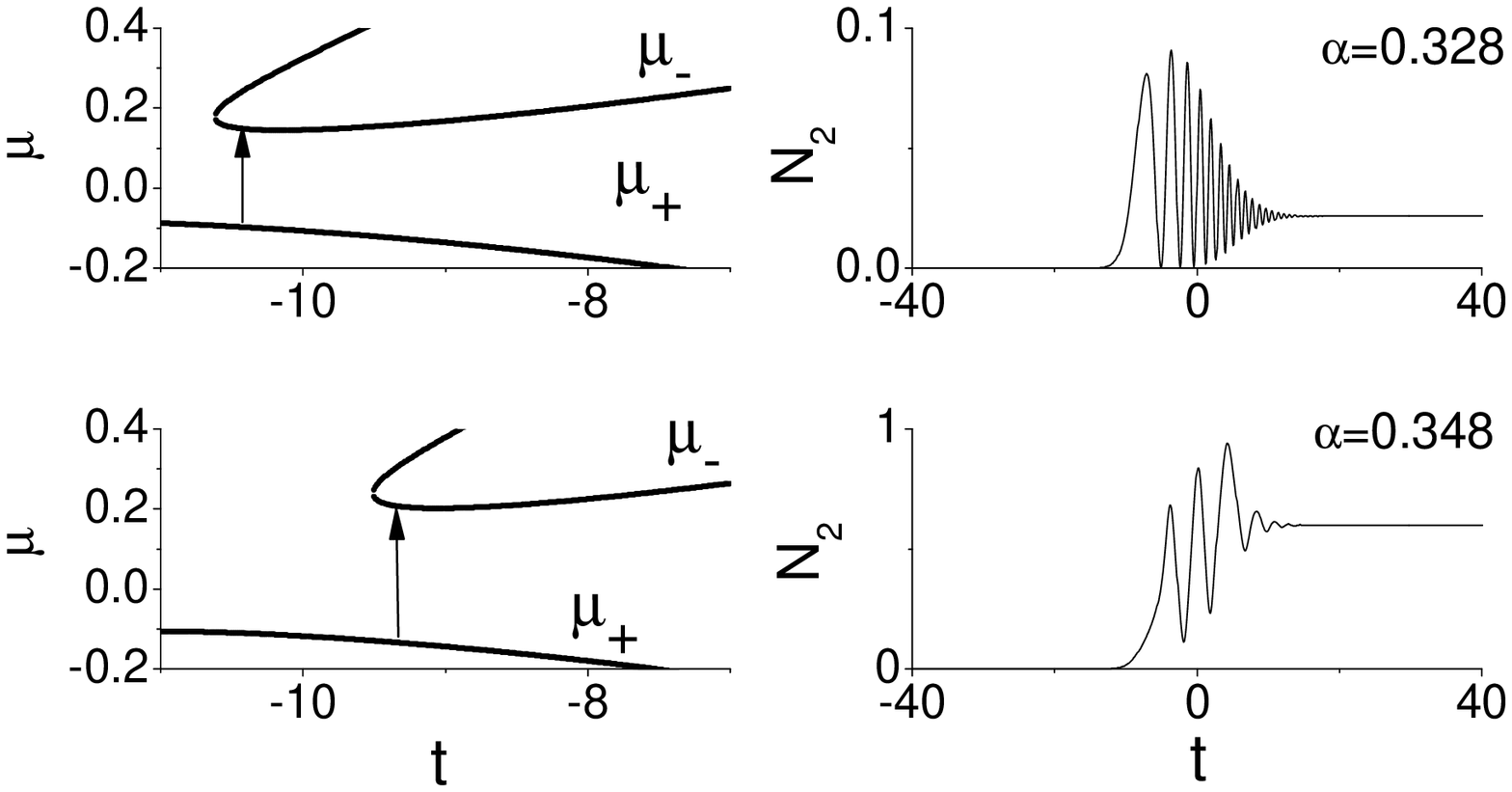}
\caption{
\label{fig:fig6_gap}
GLZ evolution of adiabatic energies $\mu_{\pm}$ (left) and
population $N_2$ (right) for the detuning rates $\alpha$=0.328 and
0.348. In all the panels $\Lambda$=4 is used.
The vertical arrows indicate the minimal energy gap between $\mu_{-}$
(upper curve) and loop edge of $\mu_{+}$ (lower curve).
}
\end{figure}

The previous analysis suggests that at $t_c \ge \Gamma$ the success of adiabatic
following is determined by the minimal gap between $\mu_+$ and $\mu_-$.
Fig. \ref{fig:fig6_gap} illustrates this point for two small detuning rates
which hinder and favor the transfer by giving  $N_2\sim$0.02  and $\sim$ 1,
respectively. This example shows that Gaussian coupling
provides a complete transfer switch by a tiny tuning of $\alpha$ and
his switch is determined by the lateral gaps between $\mu_+$ and $\mu_-$.
This property can be used for effective coherent control of BEC transport.
\begin{figure}[t]
\includegraphics[height=7.5cm,width=5.1cm,angle=-90]{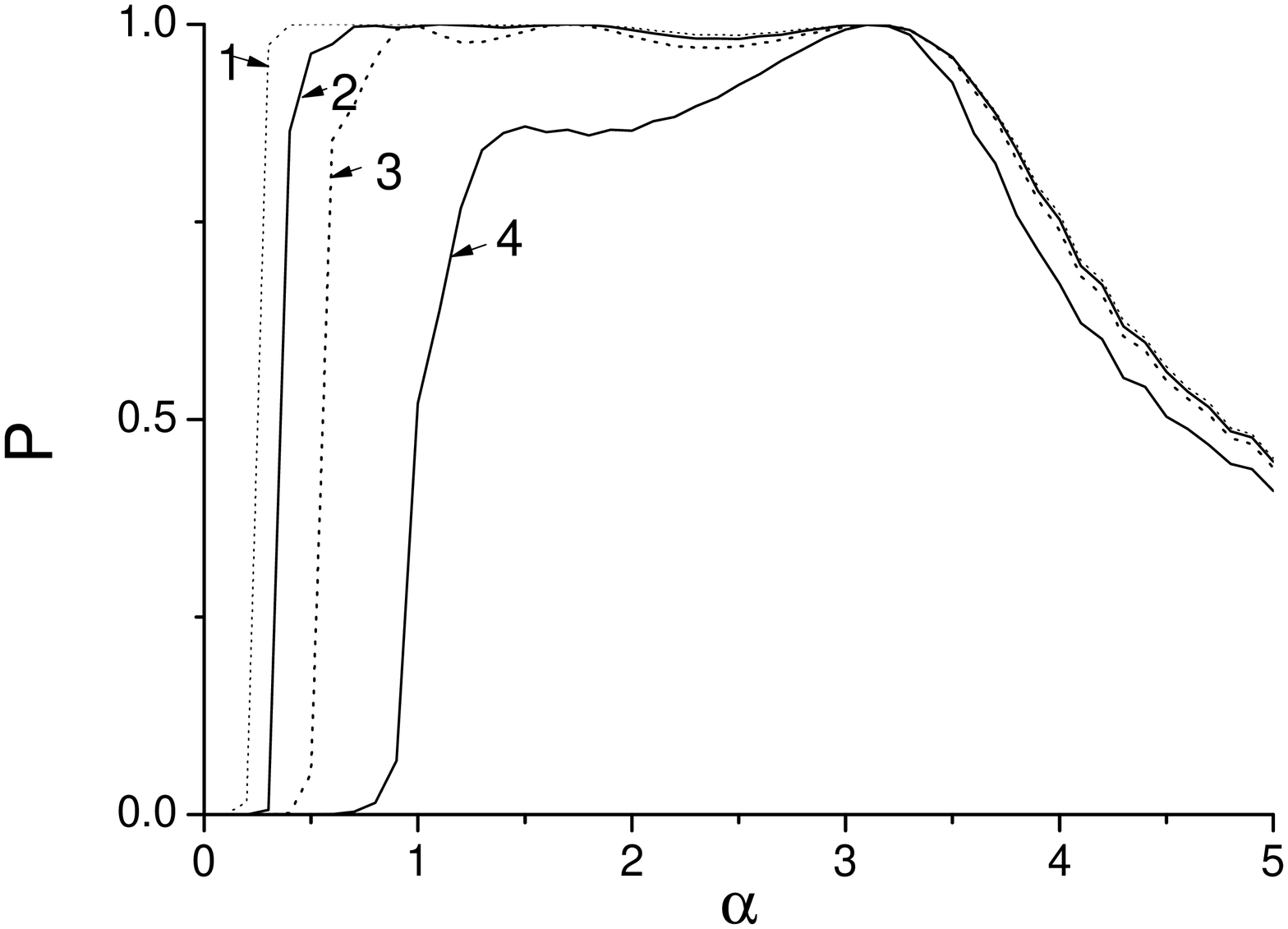}
\caption{ \label{fig:fig7_gam_dep}
The adiabatic population $P(\alpha)$ for
$\Lambda$ =4 and Gaussian widths
$\Gamma$=7.8, 5, 3.5, 2.1 (marked by 1,2,3,4, respectively). For the convenience
of comparison, the case 1 and 3 (2 and 4) are depicted by dotted (solid) curves.
}
\end{figure}
\vspace{1cm}
\begin{figure}
\includegraphics[height=7cm,width=6.2cm,angle=-90]{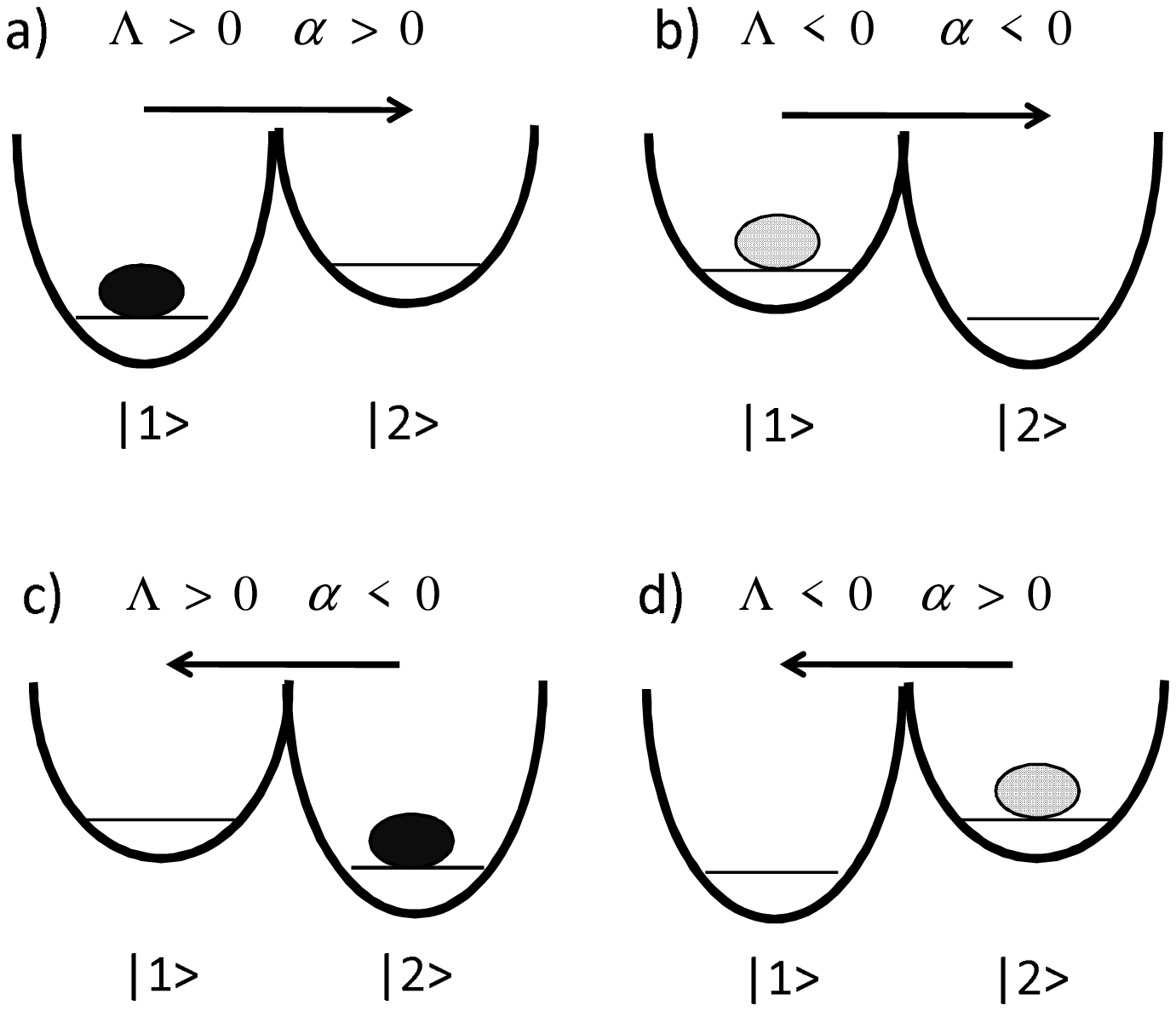}
\caption{ \label{fig:fig8_symm}
Transport schemes for repulsive ($\Lambda >0$, black dot)
and attractive
($\Lambda <0$, grey dot) BEC in directions indicated by arrows. Note relation
of the sign of $\alpha$ with initial relative positions of the well depths.
}
\end{figure}

The Gaussian coupling delivers a new control parameter $\Gamma$.
Dependence of the transport on $\Gamma$ is illustrated in
Fig. \ref{fig:fig7_gam_dep}. It is seen that increasing $\Gamma$ makes the
Gaussian coupling closer to LZ case and leads to the downshift of
the left plateau edge. Instead, decreasing $\Gamma$ upshifts the plateau edge.
Hence by changing $\Gamma$ one can
shift the switch value $\alpha_{s}$ which, following a simple estimation,
reads
\begin{equation}
\alpha_{s}\simeq \frac{\Lambda}{2 \Gamma} .
\label{alphacr}
\end{equation}
The figure shows that already at $\Gamma < 3 $ the coupling becomes too short
to support a slow adiabatic evolution and transport tends to vanish.
Note that the value of $\Gamma$ influences only the
left side of the plateau and does not affect its right side.

In the discussion above, the successful left-to-right transport of
repulsive BEC ($\Lambda >0$) was considered.
This protocol is depicted in Fig.\ref{fig:fig8_symm}a).
However, by using the symmetry (\ref{symm_1}),
the similar transport can be produced for the attractive
condensate ($\Lambda <0$) if we change the sign of the detuning rate $\alpha$
(Fig.\ref{fig:fig8_symm}b)).
Further, following the symmetry (\ref{symm_2}), the right-to-left transport
is also possible for the attractive (at $\alpha >0$) and repulsive
(at $\alpha <0$) BEC, as is shown in Fig.\ref{fig:fig8_symm}c-d).
So our protocol is quite universal.
It can be also used for other kinds of multicomponent BEC,
e.g. for two-component BEC in a single-well trap.

As was mentioned in the previous sections, the GLZ is also a
generalization of the Rosen-Zener method \cite{RZ_PR_32}
where the coupling is time-dependent
but diabatic energies are constant. Note that, unlike the GLZ,
the nonlinear RZ protocol totally blocks the adiabatic transfer
\cite{Liu_PRA_08}.

Finally note that a possible (though weak) diabatic leaking $P_{LZ}$,
appearance of the loops in nonlinear stationary spectra, and finite
rates $\alpha$ signify that the actual transport is not perfectly
adiabatic. However, being driven by adiabatic protocols, the transport
demonstrates high effectiveness and completeness in a wide range
of nonlinearity. This proves that adiabatic population transfer schemes are
indeed very robust and promising.

\section{Conclusions}

We propose a simple and effective adiabatic population
transfer protocol for the complete and irreversible transport of
Bose-Einstein condensate (BEC) in a double-well trap by using a pulsed
coupling of a Gaussian form. Being mainly based on the familiar
Landau-Zener (LZ) scheme with a constant coupling and having much in common
with that scheme, our protocol is, nevertheless, more realistic and
flexible. The protocol delivers the additional control parameter, the coupling
width $\Gamma$, which  allows to switch adiabatic transport at the detuning rate $\alpha$
beyond the LZ adiabatic limit $\alpha \to 0$, and, what is important,
provides a new transport regime when not the center but edges
of the coupling are decisive. As a result, an effective control of
BEC transport, e.g. its rapid complete switch, becomes possible.
In spite of asymmetric impact of nonlinearity, the protocol is quite
universal and can be applied to both direct and inverse transport of BEC with
both repulsive and attractive interaction. In other words, the protocol can always
be cast to enforce the nonlinearity not hamper  but favor the adiabatic
transfer. The stronger the nonlinearity, the wider the range of the detuning rate
$\alpha$ where the complete transfer is possible. The protocol is actually a
a generalization of both Landau-Zener and  Rosen-Zener schemes.

Our findings can open new opportunities
in BEC dynamics like excitation of topological states \cite{Will_PRA_00,Yuk_04},
exploration of diverse geometric phases generated in BEC transport
\cite{Ne_LP_09}, etc. The later is especially  interesting in relation to
perspectives of using geometric phases in quantum computing
\cite{Nayak_RMP_08_gp,Zhu_PRL_03_gp,Feng_PRA_07_gp}.

\begin{acknowledgments}
 The work was supported by the grants 08-0200118 (RFBR, Russia) and 684
(Universit{\'e} Paul Sabatier, Toulouse, France, 2008). We thank V.I. Yukalov
for useful discussions.
\end{acknowledgments}


\begin{thebibliography}{100}
\bibitem{collaps} 
  It is assumed that attractive BEC has sufficiently small number of atoms $N$
  to avoid the collaps. For given $U$ and $N$, it is always possible to get a
  desirable value of $\Lambda$ by tuning the coupling amplitiude $K$.
\bibitem{weak_nonlin}
  At low $\alpha$, a small nonlinearity $|\Lambda|<1$ can also result in various
  nonlinear structures (loops, double-loops, etc) and transport regimes,
  see discussion in \cite{ne_LP_10}.
\end{thebibliography}
\end{document}